\documentclass[aps,twocolumn,prb,superscriptaddress,showpacs]{revtex4}
\usepackage{amsmath,amssymb,latexsym}
\usepackage[dvips]{graphicx}

\begin{document}

\title{Near Kohn anomalies in the phonon dispersion relations
of lead chalcogenides}

\author{Ondrej Kilian}
\affiliation{Department of Astronomy, Physics of the Earth, and Meteorology,
Comenius University, Mlynska dolina F1, 842 48 Bratislava 4, Slovakia}
\affiliation{Institute for Electronics, Microelectronics, and Nanotechnology,
Dept. ISEN, CNRS-UMR 8520, B.P. 60069, 59652 Villeneuve d'Ascq Cedex, France}

\author{Guy Allan}
\affiliation{Institute for Electronics, Microelectronics, and Nanotechnology,
Dept. ISEN, CNRS-UMR 8520, B.P. 60069, 59652 Villeneuve d'Ascq Cedex, France}

\author{Ludger Wirtz}
\affiliation{Institute for Electronics, Microelectronics, and Nanotechnology,
Dept. ISEN, CNRS-UMR 8520, B.P. 60069, 59652 Villeneuve d'Ascq Cedex, France}

\date{\today}

\begin{abstract}
We present {\it ab-initio} phonon dispersion relations for the three
lead chalcogenides PbS, PbSe, and PbTe. The acoustic branches
are in very good agreement with inelastic neutron-scattering data
and calculations of the specific heat give good agreement
with experimental data.
The pronounced minimum of the transverse optical branch at $\Gamma$
due to the near-ferroelectricity of the lead chalcogenides
is qualitatively reproduced. In addition, we find a pronounced dip in 
the longitudinal optical branch at $\Gamma$.
This dip was previously explained as the effect of 
``free carriers'' (due to the presence of impurities). 
The calculations demonstrate that it persists also in the 
case of pure lead chalcogenides. We explain the dip as a "near Kohn
anomaly" which is associated with the small electronic band-gap
at the high-symmetry point L.
\end{abstract}

\maketitle

\section{Introduction}

Lead chalcogenides (PbS, PbSe, PbTe) are IV-VI narrow gap semiconductor 
compounds with rock salt crystal structure. In nanocrystalline form these 
materials manifest superior optical and electrical properties which opens 
a wide field of applications. Their small gap (280 - 410 meV at room
temperature\cite{landolt}), a large 
exciton diameter (e.g., $20nm$ in PbS and $46nm$ in PbSe \cite{wise}) and 
small effective electron and hole masses make them a good medium 
for optoelectronics, photovoltaic devices and quantum confinement 
studies~\cite{wise,wehrenberger}. An infrared diode laser was 
constructed~\cite{PbSe diod laser} based on PbSe/PbEuTe quantum dots. 
Recent studies~\cite{klimov} indicate that PbSe nanocrystals might be good
candidates for high-efficient solar cells. Because of impact 
ionization (an electron-hole pair with large energy decays into several
electron-hole pairs with lower energy), efficient carrier multiplication 
occurs.

In order to fully understand the de-excitation of hot carriers - in particular,
the ratio of radiative versus non-radiative decay channels and the mechanisms
of line broadening - it is necessary
to learn more about the electron-phonon coupling in lead chalcogenides.
While high-quality calculations of the electronic bands are available
(Ref.~\onlinecite{hummer} and references therein), the understanding
of the phonon dispersion of the lead chalcogenides is much less complete.
Experimentally, phonon dispersion relations of PbS, PbSe, and PbTe were 
obtained by inelastic neutron scattering measurements
\cite{experiment-PbTe,experiment-PbS,experiment-PbSe}. 
Simulations of the phonon dispersions have been done so far mainly on the level
of the semi-empirical shell model \cite{experiment-PbTe,experiment-PbS,
experiment-PbSe,upad,maksimenko}.
All three materials exhibit the same anomalies in the dispersion relation:
A strong softening of the TO (transverse optical) phonon branch 
around $\Gamma$ and an unexpected dip of the  LO (longitudinal optical) 
branch at $\Gamma$. 
The TO softening is due to the near-ferroelectric
character of the lead chalcogenides~\cite{jantsch} (in a truly ferroelectric
material, this mode would acquire a complex frequency, i.e., the fcc
structure would no longer be the most stable one).
Different explanations were proposed for the LO dip. 
Cowley and Dolling \cite{Cowley&Dolling} attributed the dip 
to screening by free carriers in the crystal. 
A term for free carrier doping
was consequently introduced in the recent semi-empirical phonon calculations
by Upadhyaya et al.~\cite{upad}. 
Maksimenko and Mishchenko~\cite{maksimenko} explain the LO dip 
by the dipolar pseudo-Jahn-Teller effect~\cite{pJT}. 
Recent {\it ab-initio} calculations of the phonon dispersions
of PbS, PbSe, and PbTe \cite{romero08} displayed a minimum of the 
LO mode at $\Gamma$ which (for PbSe and PbTe) turned into a strongly 
pronounced dip when spin-orbit coupling was taken into account.
The reason for the strong dip enhancement remained open.

We present in this paper a systematic {\it ab-initio} study of the three
phonon dispersion relations of the lead chalcogenides. 
The small gap and the near-ferroelectric behavior 
demand a careful choice of calculation parameters such as sampling grid and 
pseudopotentials. Our calculations reproduce quantitatively the acoustic modes.
The anomalies of LO and TO modes are qualitatively reproduced 
in the calculations (since they are strongly temperature dependent 
\cite{maksimenko}, anharmonic effects would have to be taken into account
in order to quantitatively reproduce the measurements which were
performed at room temperature). Our calculations show
that a pronounced LO dip is present in the pure materials
even without free carrier doping. We explain this dip in analogy
to the Kohn anomalies \cite{kohn} that occur in the semi-metal
graphene \cite{piscanec04}.
Furthermore, we demonstrate that {\it ab-initio} calculations can
reproduce very well the specific heat of PbS and compare with
recent calculations \cite{cardona07,romero08} that include
the effect of spin-orbit effects.

In section \ref{details} we summarize the computation method
and give the details of the calculations. In section \ref{dispersion}
we discuss the dispersion relations of the three lead chalcogenides.
Section \ref{cvsec} presents data on the specific heat in comparison
with experimental data. In the appendix, we show that the electronic
k-point sampling must be dense enough around the high-symmetry point
L in order to reproduce the LO dip at $\Gamma$.

\section{Computational Details}
\label{details}

In the harmonic approximation,
the phonon frequencies (as a function of the wave-vector ${\bf q}$) are 
obtained from the equation
\begin{equation}
\det{\lvert\frac{1}{\sqrt{M_sM_t}}\widetilde{C}_{st}^{\alpha
\beta}({\bf q})-\omega^{2}({\bf q})\rvert}=0.
\end{equation}
The dynamical matrix $\widetilde{C}_{st}^{\alpha
\beta}({\bf q})$ corresponds to the force on atom 
$t$ in direction $\beta$ linearly induced by a displacement of atom $s$ in
direction $\alpha$. We calculate it with density functional perturbation
theory (DFPT)~\cite{DFPT,baroni} as implemented in the code {\tt ABINIT}
\cite{abinit}. We use the local density approximation (LDA) for the
exchange-correlation functional \cite{LDA}. The wave-functions are
expanded in plane waves. Core electrons are replaced by pseudopotentials. 

We found that the phonon frequencies
are very sensitive to the choice of the lead pseudopotential. In particular,
it is important to include the lead 5d semi-core electrons as valence
electrons in the calculation.
(This is different from the case of pure lead, where the 5d electrons 
do not alter the phonon frequencies significantly \cite{verstraete}).
The reason lies in the strongly ionic character of the lead chalcogenides:
the lead atoms tend to transfer the 6p valence electrons to the anions.
It is then the overlap of the remaining 6s and 5d electrons of lead with the
3p orbitals of the anion that determines the covalent part of the PbX
bond (where X stands for the anion S, Se, or Te, respectively).
Since the 5d orbitals contribute to this bonding, their density
should be calculated explicitly and not be substituted by a pseudopotential. 
We tested different Troullier-Martins pseudopotentials created with the 
FHI pseudopotential generation code \cite{fhi}.
We verified that the corresponding FHI potential for lead and the chalcogens
from the {\tt ABINIT} web-page 
yielded converged results for the phonon frequencies. The plane-wave
energy cutoff is 40 Ha.

It has been observed for ferroelectric materials that {\it ab-initio} phonon 
calculations give better agreement with experimental data if they are
performed at the experimental lattice constant, rather than the 
lattice constant obtained by total energy minimization 
(see Ref.~\onlinecite{resta} and references therein).
The reason is that the LDA tends to underestimate the lattice constant
and in (near) ferroelectric materials even a small underestimation
of 1\% strongly influences the ferroelectric instability.
Our phonon calculations are performed using the experimental lattice
constants at 300K. The experimental lattice constant are given
in Table~\ref{latconsttable} together with the values of the
optimized lattice constants.
Since anharmonic effects are neglected (and very difficult to
include on an {\it ab-initio} level \cite{bonini}), 
we do not expect to fully reproduce the temperature dependence
of the phonon dispersions.

\begin{table}
\begin{tabular}{|l|c|c|c|} \hline
 & DFT-LDA & Exp. (30 K) & Exp. (300 K) \\
\hline
PbS  & 5.810 {\AA} & 5.909  {\AA} & 5.936  {\AA}\\
PbSe & 6.012  {\AA}& 6.098  {\AA} & 6.124  {\AA}\\
PbTe & 6.318  {\AA}& 6.428 {\AA}  & 6.462  {\AA}\\
\hline
\end{tabular}
\caption{Calculated lattice constants in comparison
with experimental lattice constants \cite{landolt} at 30 K and at 300 K.}
\label{latconsttable}
\end{table}

The influence of the electronic k-point sampling on the phonon
dispersion is discussed in the appendix. For converged results,
we used a (7,3) nested grid, i.e., a
7$\times$7$\times$7 (shifted) Monkhorst-Pack shifted k-point sampling with an
additional 3$\times$3$\times$3 sampling of the volume element
around the high symmetry point L.
In order to obtain the phonons at arbitrary phonon wave vector ${\bf q}$,
the dynamical matrix $\widetilde{C}_{st}^{\alpha \beta}({\bf q})$ is
calculated on a 8$\times$8$\times$8 mesh and then Fourier-interpolated
for arbitrary ${\bf q}$. In order to properly reproduce the LO dip
around $\Gamma$, we calculated the dynamical matrix explicitly
for a set of ${\bf q}$-points along the high-symmetry
lines $\Delta$, $\Sigma$, $\Lambda$ close to the $\Gamma$ point.

\section{Phonon dispersion relations}
\label{dispersion}
\begin{figure}[htpb]
 \centering
   \includegraphics[draft=false,keepaspectratio=true,clip,
                    width=1.0\linewidth]
                    {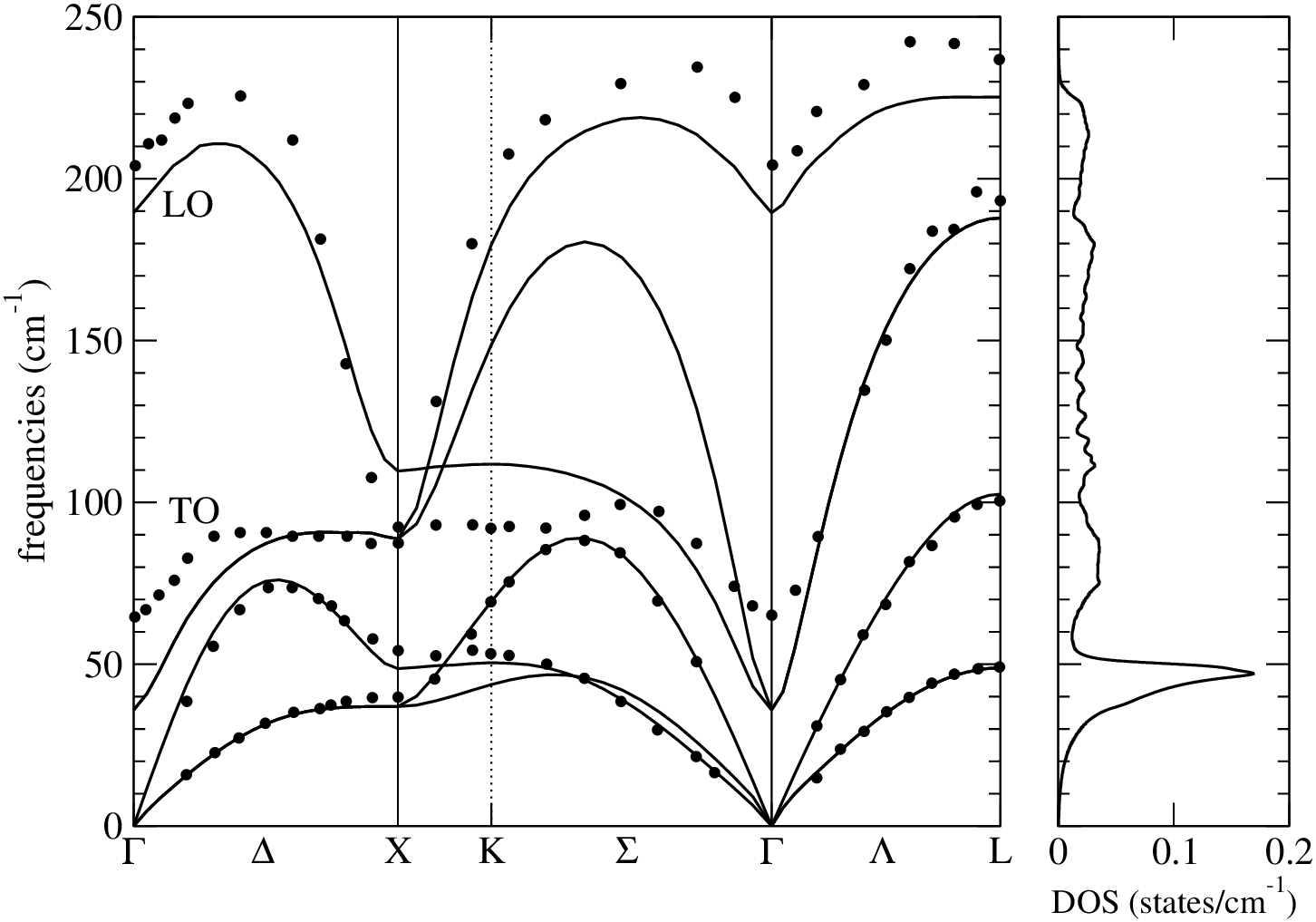}
\caption[FIG1]{Left panel: Calculated phonon dispersion relation of lead 
sulfide (lines) in comparison with experimental data~\cite{experiment-PbS} 
(dots). Right panel: phonon density of states.}
\label{pbsdisdos}
\end{figure}

\begin{figure}[htpb]
 \centering
   \includegraphics[draft=false,keepaspectratio=true,clip,
                    width=1.0\linewidth]
                    {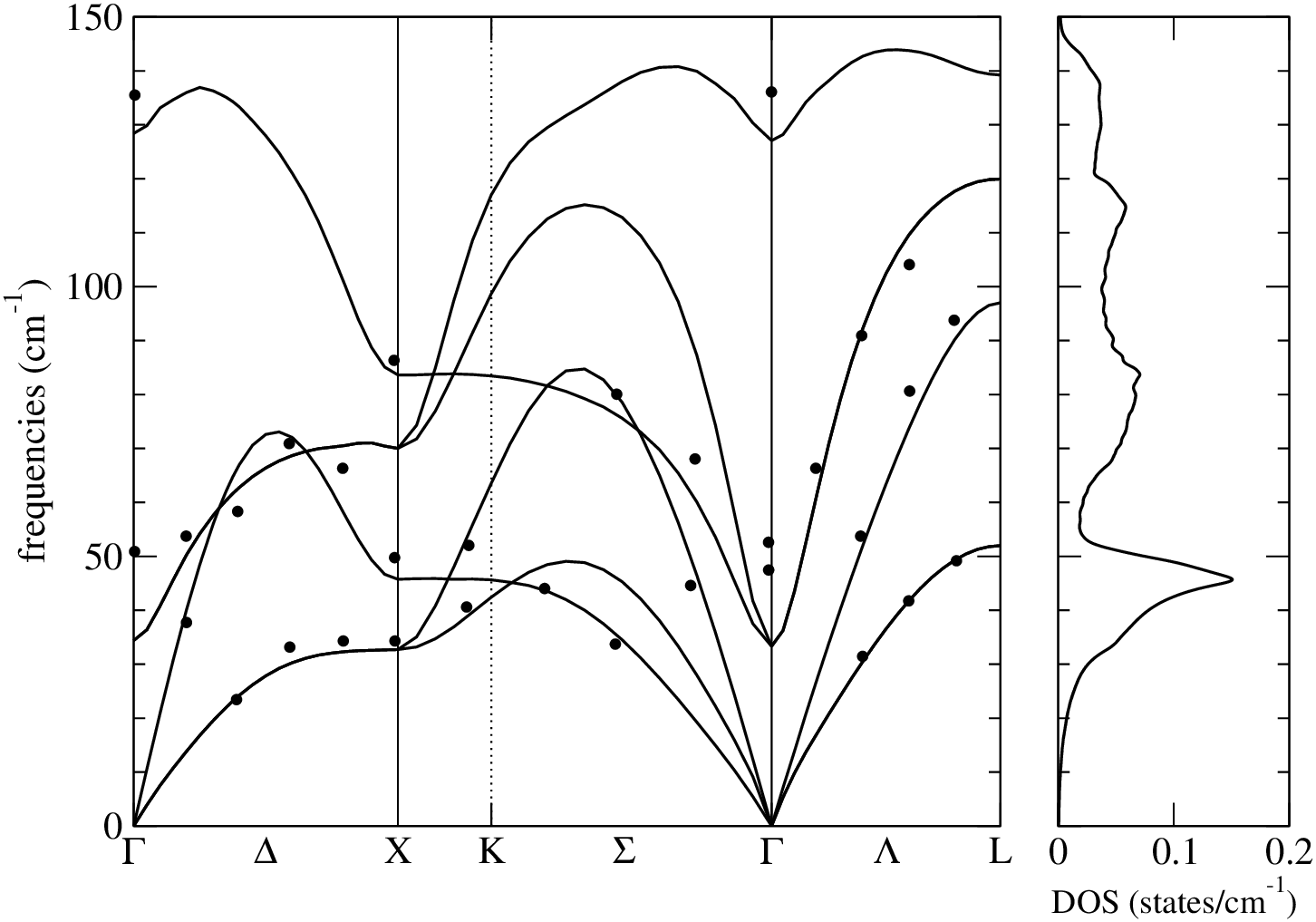}
\caption[FIG2]{Calculated phonon dispersion relation of lead selenite (lines) 
in comparison with experimental data~\cite{experiment-PbSe} (dots).}
\label{pbsedisdos}
\end{figure}

\begin{figure}[htpb]
 \centering
   \includegraphics[draft=false,keepaspectratio=true,clip,%
                    width=1\linewidth]
                    {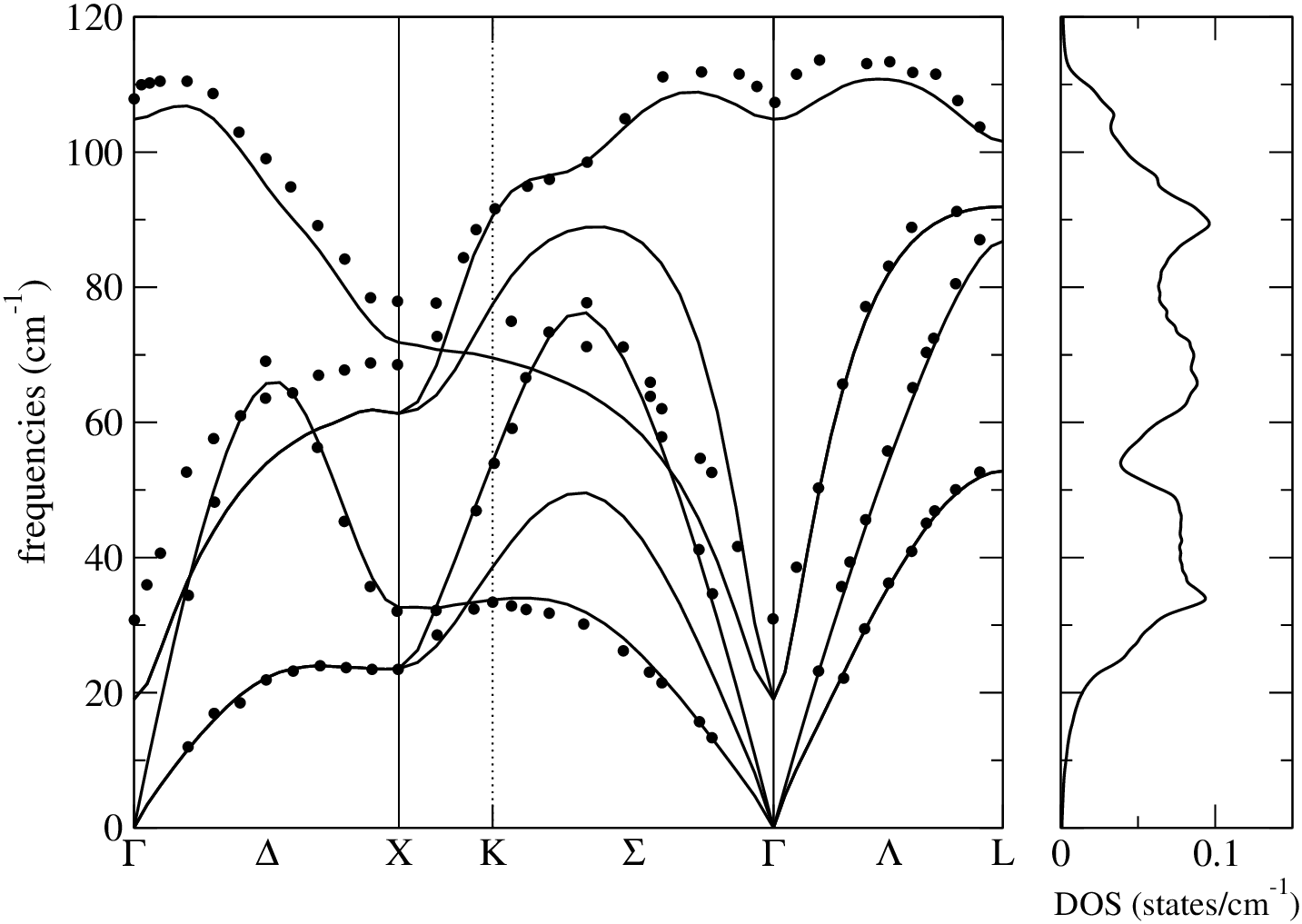}
\caption[FIG3]{Calculated phonon dispersion relation of lead telluride (lines) 
in comparison with experimental data~\cite{experiment-PbTe} (dots).}
\label{pbtedisdos}
\end{figure}

In Figs.~\ref{pbsdisdos},\ref{pbsedisdos}, and \ref{pbtedisdos},
we present our calculated dispersion relations for PbS, PbSe, and PbTe,
respectively, and compare with experimental data from inelastic
neutron scattering. We also show the calculated phonon densities of states
which will be needed for the calculation of the specific heat.
In all three cases, we obtain excellent agreement between theory and experiment
for the three acoustic modes. Since the mass of Pb atoms
is much larger than the masses of S, Se, and Te,
the acoustic branches correspond almost exclusively to vibrations 
of lead ions. Consequently, the acoustic mode dispersion is quantitatively
very similar in the three lead chalcogenides. The main difference
is a small frequency lowering that corresponds to increasing lattice
constant in the series PbS, PbSe, PbTe.
The three optical branches correspond to vibrations of the anions.
Their maximum frequency scales as $1/\sqrt{M_X}$, where X denotes
the anion species. For PbS, the optical modes are so high
in energy that they do not intersect with the acoustic modes. 
For PbSe and PbTe some crossings occur.

In all three cases, we observe a remarkably large LO/TO splitting
at $\Gamma$. For cubic systems, the splitting is described
by the relation \cite{baroni}
\begin{equation}
\omega_{LO}^2 - \omega_{TO}^2 \propto (Z^*)^2/\epsilon,
\label{loto}
\end{equation}
where $Z^*$ is the effective charge and
$\epsilon$ is the dielectric constant.
Since the effective charges are very large
($\pm 4.5$, $\pm 4.9$, and $\pm 6.1$ for PbS, PbSe, and PbTe,
respectively), the LO/TO splitting is strongly pronounced
and the TO mode at $\Gamma$ has a very low frequency. This frequency 
depends sensitively on the lattice constant. E.g., for PbSe, an artificial
increase of the lattice constant by 0.1 {\AA} leads to a softening
of the TO mode frequency below 0, i.e., towards imaginary values
which means that the fcc phase would no longer be the stable one.
This is a clear manifestation of the near-ferroelectric character
of the lead chalcogenides.

All three phonon dispersions exhibit a significant frequency drop
of the LO mode at $\Gamma$. Cowley and Dolling~\cite{Cowley&Dolling} 
proposed that this phenomenon is caused by the screening of the 
macroscopic electric field accompanying the LO mode by free
carriers (which may have their origin by doping from impurities). 
We have done our calculations without the presence of additional free 
carriers, yet we have reproduced the LO anomaly.
Free carrier doping can contribute
to the LO dip but seems not to be its primary cause.
Our calculations are in accordance with
the theory of Maksimenko and Mischenko~\cite{maksimenko} who
explained the LO anomaly as due to a strong electron-phonon
interaction of pseudo-Jahn-Teller type (in absence of free
carrier doping). 

We explain this dip as a "near Kohn anomaly". The notion of a
Kohn anomaly \cite{kohn} is known from metallic systems: the vibrations
of the ionic cores are partially screened by the surrounding electron gas.
The screening can be strongly enhanced for vibrations with a wave-vector
${\bf q}$ that connects two points on the Fermi surface. The enhanced
screening then leads to a dip in the phonon dispersion at those values
of ${\bf q}$. Recently, two Kohn anomalies were found in the
semi-metal (or "zero gap semiconductor") graphene 
\cite{piscanec04}. The band structure in the first Brillouin zone
of graphene displays two conical intersections (linear crossings)
of the $\pi$ and $\pi^*$ bands at the Fermi level. 
The Fermi surface is thus reduced to two points and Kohn anomalies
can be found at $\Gamma$ (${\bf q}=0$) and at K.

\begin{figure}[htpb]
 \centering
   \includegraphics[draft=false,keepaspectratio=true,clip,%
                    width=1\linewidth]
                    {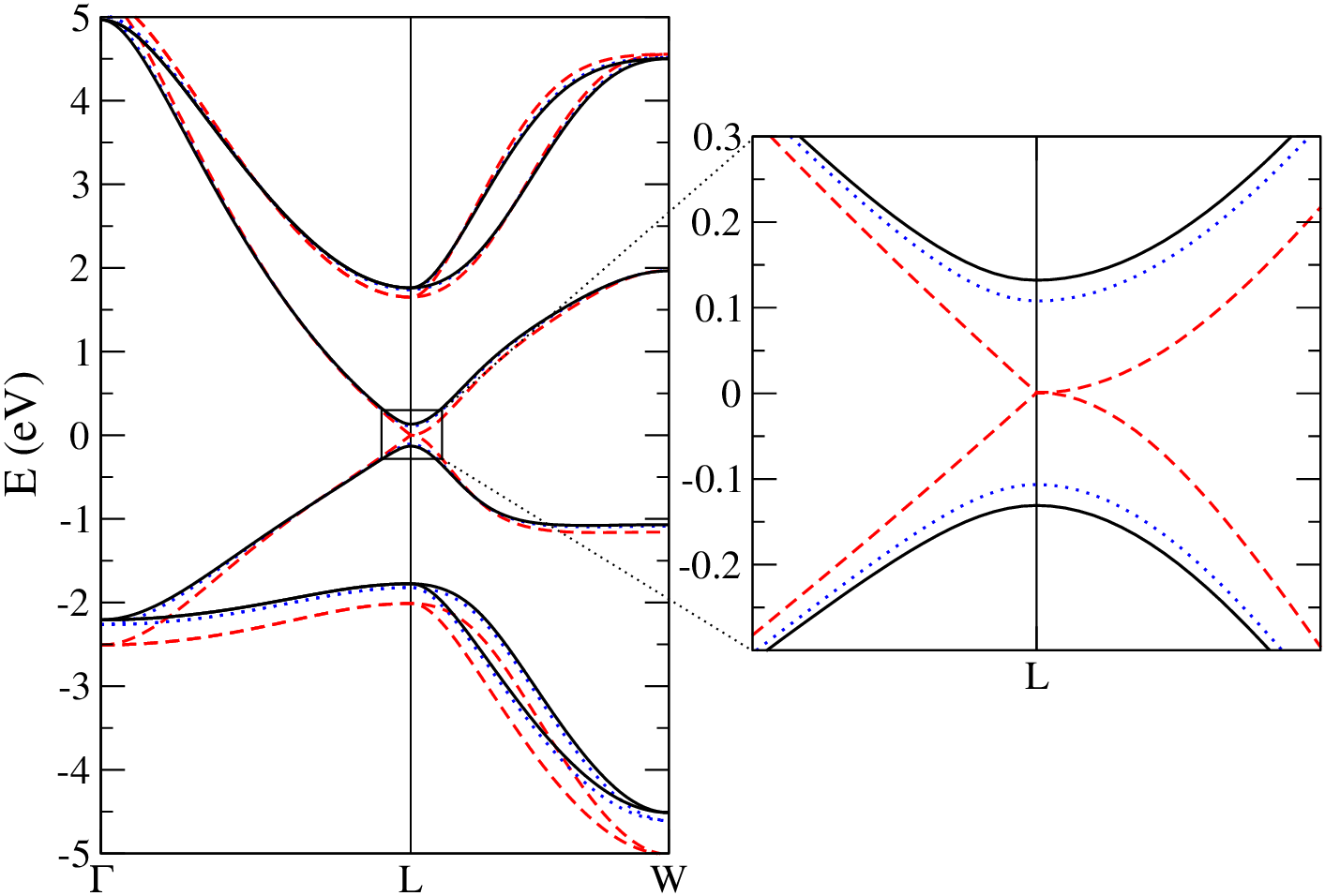}
\caption{(Color online) Left panel: Electronic band-structure (DFT-LDA
without spin-orbit coupling) of PbS for three
different values of the lattice constant $a$. Black solid lines: 
experimental lattice constant at 300 K; blue dotted lines: 
experimental lattice constant at 30 K;
red dashed lines: $a=5.801$ {\AA} (squeezed lattice constant
which reduces the gap to zero). The Fermi level is at 0 eV.
A zoom for the region around the direct gap at L is presented
in the right panel.}
\label{pbsbandstruc}
\end{figure}

The lead chalcogenides are semiconductors. Thus there
are - a priori - no Kohn anomalies in their phonon dispersions.
However, the direct band-gap at the high-symmetry point L is small
and, by compressing the lattice, it can be brought down to zero.
The situation is demonstrated for the case of PbS in Fig.~\ref{pbsbandstruc}.
We show the band-structure along the high-symmetry lines
$\Gamma$ $\rightarrow$ L $\rightarrow$ W for three different values
of the lattice constant. The highest valence band around L
is composed of S 3$p$ orbitals. The lowest conduction band 
has a Pb 6$p$ character with an admixture of S 4$s$ orbitals.
When calculated with the experimental lattice constant at 300 K, 
the LDA-DFT band gap is 267 meV. It decreases to the value of 
216 meV for the 30 K experimental lattice constant.
Artificially decreasing the lattice constant further reduces
the band-gap at L. Eventually, for a value of the lattice
constant $a = 5.801$ {\AA}, the band gap becomes zero 
with a linear crossing of the bands in the direction L $\rightarrow$ $\Gamma$ 
and a parabolic dispersion in the direction L $\rightarrow$ W
(see right panel of Fig.~\ref{pbsbandstruc}).
This situation is now quite analog to the situation in graphene:
the system is semimetallic and the Fermi surface is pointlike,
located at the high symmetry points L. For the corresponding
phonon dispersion relation one can expect a Kohn anomaly at $\Gamma$
and at X (the wave-vector difference between to different points L
corresponds to either $\Gamma$ or X).
We note in passing that a further reduction of the lattice constants re-opens
the gap at L. However, the character of valence and conduction band
is inverted and a real crossing of valence and conduction bands
along the line L $\rightarrow$ W shows up.

\begin{figure}[htpb]
 \centering
   \includegraphics[draft=false,keepaspectratio=true,clip,%
                    width=1\linewidth]
                    {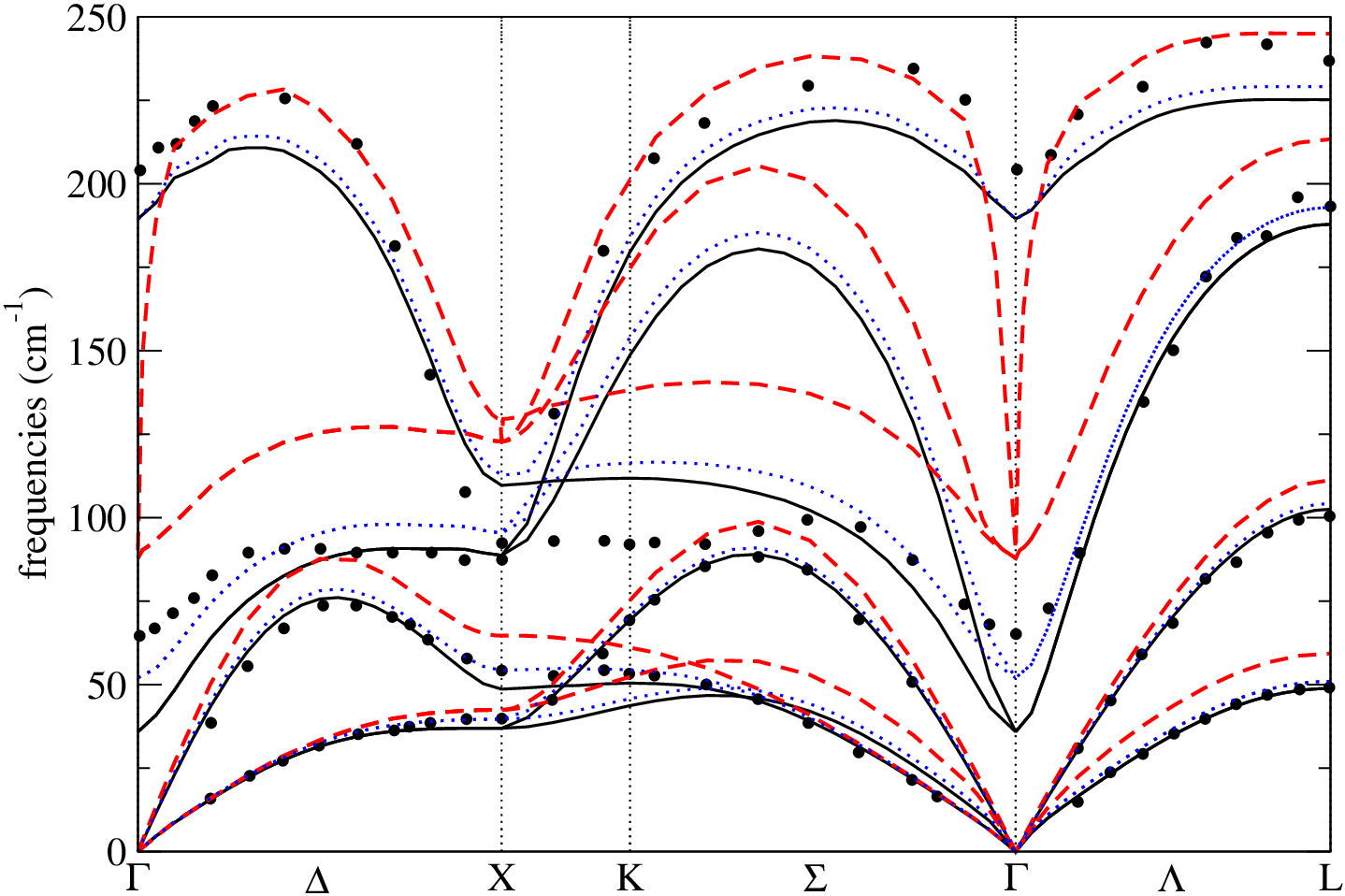}
\caption{(Color online) Calculated phonon dispersion of PbS for three
different values of the lattice constant $a$. Black solid lines: 
experimental lattice constant at 300 K; blue dotted lines: 
experimental lattice constant at 30 K; 
red dashed lines: $a=5.801$ {\AA}\ (squeezed lattice constant
which reduces the gap to zero). Circles: experimental data
\cite{experiment-PbS}.}
\label{pbsshrink}
\end{figure}

Fig.~\ref{pbsshrink} presents the phonon dispersion relations of PbS
for the three different values of the lattice constant.
As expected, shrinking the lattice constant leads to a stiffening of the
bonds and thus to an increase of the phonon frequencies.
In particular, the TO mode is strongly affected because 
shrinking the lattice constant makes the system ``less ferroelectric''.
The only mode which softens is the highest optical branch
at $\Gamma$ and at X where we observe to very sharp dips
which correspond to the Kohn anomalies predicted above.
Furthermore, LO and TO mode at $\Gamma$ are degenerate for
the squeezed lattice: $\epsilon$ tends to infinity
in the limit of the semi-metallic squeezed lattice and the LO/TO splitting 
tends to zero according to Eq.~(\ref{loto}).
Opening the electronic gap by enlarging the lattice constant to its
original value reinstalls the LO/TO splitting. However, a remnant of the 
Kohn anomaly remains visible in the form of a strong dip of the 
LO mode at $\Gamma$. For this reason, we interpret the LO dip as a
"near Kohn anomaly".

\begin{table}
\begin{tabular}{|l|c|c|} \hline
 & $E^{gap}$ (calc.) & $E^{gap}$ (exp.) \\
\hline
PbS & 267 meV & 410 meV \\
PbSe & 244 meV & 280 meV \\
PbTe & 644 meV & 310 meV \\
\hline
\end{tabular}
\caption{Values of the band-gaps of the lead chalcogenides.
Calculations are performed on the DFT-LDA level with the experimental
lattice constant at 300K. Experimental data at 300K (Ref.~\onlinecite{landolt}).}
\label{gaptable}
\end{table}

We note that the LO dip is considerably less pronounced in our calculated
phonon dispersion of PbTe. This is due to the strong overestimation
of the gap of this material \cite{gapnote} (see table \ref{gaptable} 
for our values obtained on the DFT-LDA level and a comparison with the 
experimental values). At this point, we have to discuss if the inclusion
of spin-orbit coupling (SOC) leads to an improvement
of the phonon-calculations. It has been shown by Hummer et al.~\cite{hummer}
that SOC strongly reduces the electronic gap for the three lead chalcogenides.
E.g., for PbTe, the gap is reduced to 60 meV.
The inclusion of electron-correlation effects, e.g., on the level
of the GW-approximation, is needed in order to ``re-open'' the gap
and to obtain values in good agreement with the experimental 
gaps \cite{hummer}. For the calculation of the LO-dip this means that
inclusion of SOC does not necessarily yield better results
(unless electron-correlation effects are properly
taken care of at the same time): the underestimation of the gap leads 
to an overestimation of the LO dip. Such an effect can be seen, e.g., in the
PbTe phonon dispersion of Romero et al.\ (Fig.~3 of Ref.~\onlinecite{romero08}).

We note that a recent {\it ab-initio} calculation of Cardona
et al.~\cite{cardona07,romero08} yielded considerably higher phonon frequencies 
for PbS than the experimental data. This is due to two reasons:
they use the optimized lattice constant $a = 5.808 {\AA}$ which
underestimates the experimental lattice constant at 30K by 1.7\%
and the room temperature lattice constant by 2.2\%.
Furthermore, their use of a pseudo-potential with the lead 5d electrons in 
the core may lead to different phonon frequencies.
Including SOC, leads to a general softening of the phonons and improves
the agreement with experimental data \cite{romero08} 
(except for the LO dip at $\Gamma$). Our calculations with the experimental
lattice constant at 300 K are in very good agreement with the
acoustic branches of the experimental dispersions. Inclusion of SOC 
would probably lead to less agreement. 

The optical phonons are in worse quantitative agreement with
the experimental data points than the acoustic modes.
In particular, for the TO mode around $\Gamma$, strong deviations occur.
This is not surprising, since our calculations take into account
temperature effects only through the choice of the room temperature
lattice constant while phonon renormalization through phonon-phonon
interaction~\cite{bonini} is neglected within the harmonic
approximation. Due to the near-ferroelectricity, the renormalization
of the TO mode as a function of temperature will be particularly strong. 
E.g., the model calculation by Maksimenko and Mischenko~\cite{maksimenko} 
predicts for PbTe that the TO mode at room temperature stiffens by 
about 15 cm$^{-1}$ with respect to its value at 4K.

\section{Specific heat}
\label{cvsec}
Another test of the quality of our {\it ab-initio} phonon
calculations is the comparison with available experimental data
for the specific heat, $c_v$, of PbS, PbSe, and PbTe 
\cite{parkinson,bevolo,cardona07,romero08}.
The specific heat depends on the phonon densities of states, $D(\omega)$,
(right panels of Figs.~\ref{pbsdisdos},\ref{pbsedisdos},\ref{pbtedisdos}).
We calculate it numerically through the formula
\begin{equation}
c_v = N_A k \int_0^\infty \frac{ \left(\frac{\hbar\omega}{kT}\right)^2
e^{\frac{\hbar\omega}{kT}}}{\left(e^{\frac{\hbar\omega}{kT}}-1\right)^2} 
D(\omega) d\omega,
\label{cveq}
\end{equation}
where $k$ is the Boltzmann constant and $N_A$ is the Avogadro constant.
Note that $D(\omega)$ is normalized to 6, i.e., the number of phonon
branches.

\begin{figure}[htpb]
 \centering
   \includegraphics[draft=false,keepaspectratio=true,clip,%
                    width=1\linewidth]
                    {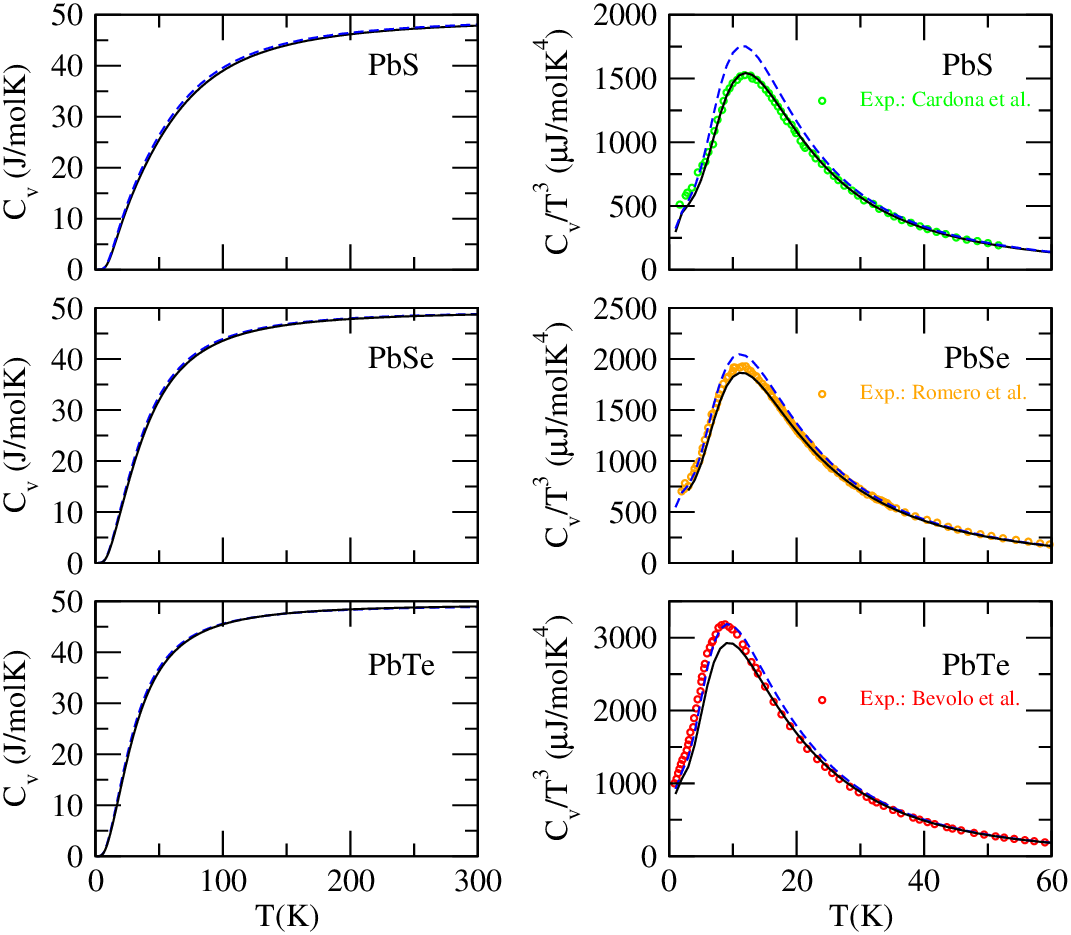}
\caption{(Color online) Left panels: Specific heat of the three lead 
chalcogenides as a function of temperature.
Right panels: Specific head divided by $T^3$ in the low temperature
regime. Results from {\it ab-initio} phonon calculations
using the respective lattice constants at 30K (solid line) and at
300K (dashed lines). Symbols: Experimental data from Refs.~\onlinecite{cardona07} (Cardona et al.), ~\onlinecite{romero08} (Romero et al.) and 
\onlinecite{bevolo} (Bevolo et al.).}
\label{cvfig}
\end{figure}

The resulting specific heat as a function of temperature is 
plotted in the left panels of Fig.~\ref{cvfig}. All three curves
display the typical convergence towards the Petit and Dulong value
$6 N_A k = 49.9 \mbox{J/molK}$ for a material with two atoms in
the primitive cell. Following the discussion of Cardona et al.\
in Refs.~\onlinecite{cardona07,romero08}, we also display $c_v/T^3$ in the
low temperature regime (right panels of Fig.~\ref{cvfig}).
All three curves display a maximum between 8 and 12 K.

The agreement between the experimental and theoretical height
of the maximum of $c_v/T^3$
was used in Ref.~\onlinecite{cardona07} as a critical test for
the quality of the {\it ab-initio} phonon calculations.
They obtained a maximum height of the {\it ab-initio} curve
at 1160 $\mu$J/molK$^4$ while the experimental height
is at 1520 $\mu$J/molK$^4$. The deviation was tentatively
assigned to the absence of SOC in the
calculations. Recently, it was shown for elemental bismuth 
\cite{diaz} and antimony \cite{serrano}
that inclusion of SOC in the phonon calculations
leads to a lowering of the acoustic modes and thus to an
increase of the maximum $c_v/T^3$.
Also for elemental lead, a lowering of the acoustic
modes through the inclusion of s-o coupling has been observed \cite{verstraete}.

Since our dispersion relations have been calculated with the room 
temperature lattice constants, we expect that we underestimate
the frequencies of the acoustic phonons at very low temperature
(where the lattice constant shrinks and the inter-atomic
force constants stiffen). Consequently, our computed $c_v/T^3$ 
for PbS should overestimate the measured one. Fig.~\ref{cvfig} (blue dashed
line) shows that this is indeed the case: we obtain the maximum $c_v/T^3$
at 1750 $\mu$J/molK$^4$. For a better assessment of the specific heat
at low temperature, we have repeated the calculation of the phonon
dispersion and the DOS using the PbS lattice constant at 30K
 (see Table~\ref{latconsttable}). The resulting $c_v/T^3$ (solid
black line in Fig.~\ref{cvfig}) is in excellent agreement with the measured
data. For PbSe, the agreement with experiment is also fairly good.
However, in the case of PbTe, the specific heat calculated with the
phonons at the low temperature lattice constant is somewhat lower
than the experimental data. 
Consequently, spin-orbit coupling (which has a stronger effect in PbTe
than in PbSe and PbS \cite{romero08}) might be needed to yield good 
agreement with the experimental data. Further calculations of the 
phonon DOS including SOC effects (and the 5$d$ electrons of Pb in the valence)
are needed to resolve this issue.

\section{Conclusion}

We have calculated the phonon dispersion relations for lead
chalcogenides. Strict convergence parameters (concerning
the lead pseudopotential and the k-point sampling)
enabled us to obtain good agreement with experimental dispersion
relations and measurements of the specific heat. The acoustic phonon 
modes are reproduced almost exactly. The pronounced dip of the LO 
mode at $\Gamma$ is related to the narrow band gap and can be understood 
as a near Kohn anomaly. 
This work provides the starting point for the investigation of
electron-phonon coupling in nanocrystals of lead-chalcogenides.

\section*{Acknowledgments}
We would like to thank C. Delerue, J. Serrano, M. Cardona, A. Romero,
X. Gonze, and M. Verstraete for 
stimulating discussions. Funding was provided by the French National Research 
Agency through project ANR PJC05\_6741. Calculations were done at the IDRIS
supercomputing center, Orsay (Proj. No. 081827).

\appendix

\section{Influence of the k-point sampling on the LO-mode frequency at $\Gamma$}
For most semiconductors, a 4$\times$4$\times$4 or 6$\times$6$\times$6 
(shifted) Monkhorst-Pack k-point sampling
\cite{Monkhorst&Pack} of the electronic structure is sufficient to reproduce 
the phonon dispersion, including the LO/TO splitting at $\Gamma$ for 
polar materials.
In this appendix, we show that for lead chalcogenides a much higher 
sampling is needed to
properly reproduce the LO mode dispersion around $\Gamma$.
\begin{figure}[htpb]
 \centering
   \includegraphics[draft=false,keepaspectratio=true,clip,
                    width=1.0\linewidth]
                    {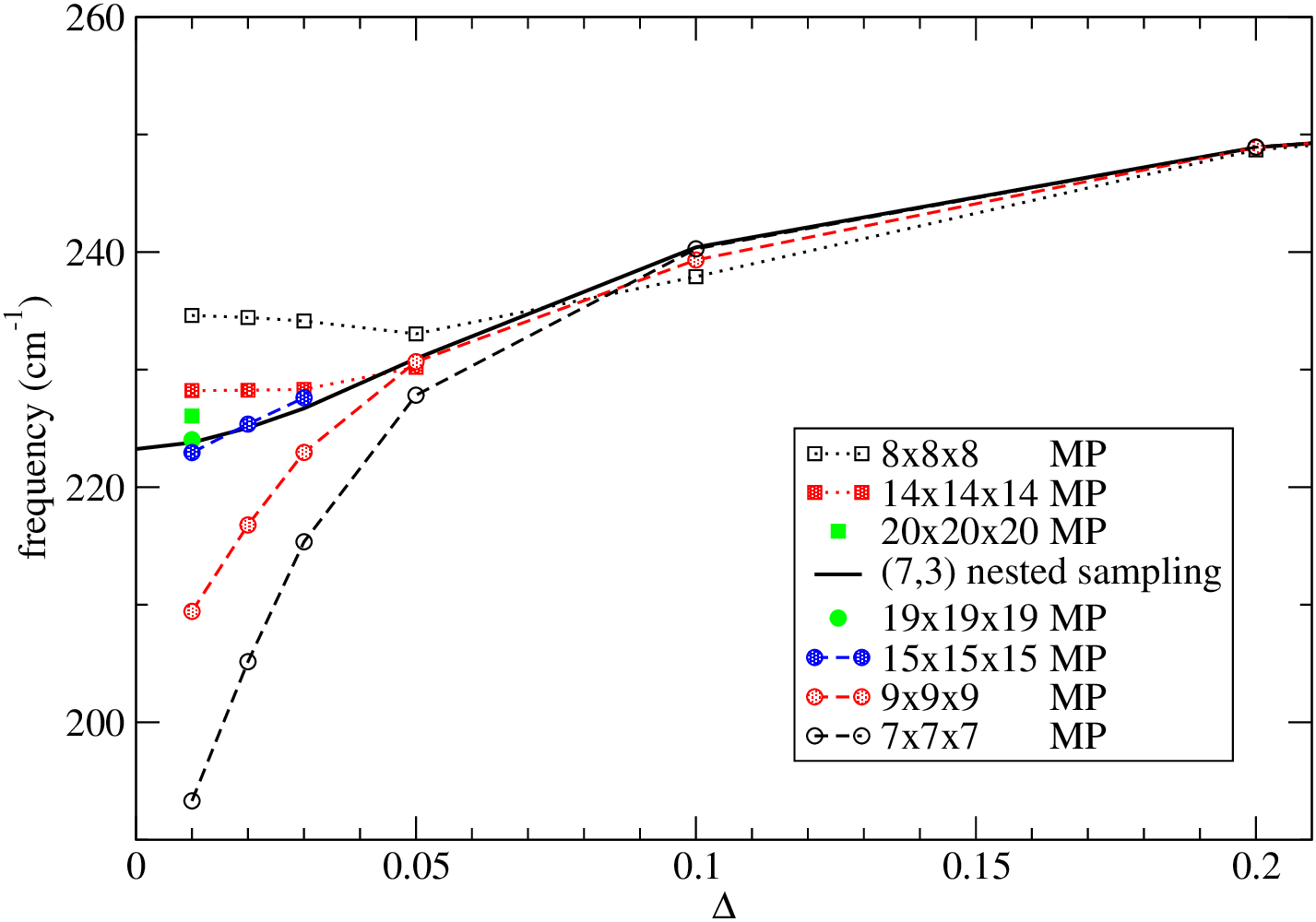}
\caption{(Color online) LO-mode dispersion in PbS for different Monkhorst-Pack 
(MP) samplings and a nested-sampling (see text). The dispersion is plotted
along the line $\Delta$ ($\Gamma \rightarrow \mbox{X}$) with 0 corresponding 
to $\Gamma$ and 0.5 corresponding to X.}
\label{samplingfig}
\end{figure}
Fig.~\ref{samplingfig} shows the LO mode dispersion close to $\Gamma$
for different n$\times$n$\times$n Monkhorst-Pack
samplings.
We used here a Troullier-Martins pseudopotential with the 5d electrons
in the core (from the ABINIT web-page). This allowed us to go to very
high k-point samplings. The phonon frequencies marked by the symbols
have been obtained by directly calculating the dynamical matrix for the
corresponding phonon wave-vector (avoiding interpolation of the
dynamical matrix).
Obviously, it makes a big difference if $n$ is even or odd.
For $n$ even, we obtain higher frequencies and for $n$ odd, we obtain
lower frequencies than in the limit $n \rightarrow \infty$.
The difference between the 7$\times$7$\times$7 and the 8$\times$8$\times$8
sampling amounts to more than 40 cm$^{-1}$ for the LO mode at $\Gamma$!
(For the TO mode - not shown here - the corresponding difference
is less than 3 cm$^{-1}$.) Even between the 19$\times$19$\times$19
sampling and the  20$\times$20$\times$20 sampling, there remains a difference
of 2 cm$^{-1}$ for the LO mode (while the TO mode is converged to within
0.03 cm$^{-1}$). 

The origin of the even-odd discrepancy for the different k-point samplings lies
in the electronic structure of the lead chalcogenides which all have
a very small direct gap at the high-symmetry point L.
The small gap is (among other factors) responsible for the very
high dielectric constants ($\epsilon > 20$) of the lead chalcogenides.
The point L is included in the samplings when n is odd, but not
when n is even. We noticed that response-function calculations with 
odd samplings tend to strongly overestimate $\epsilon$ while the even samplings 
underestimate it. The link between the dielectric screening
and the LO/TO splitting is given by Eq.~\ref{loto}.
A variation of $\epsilon$ due to insufficient k-point
sampling will strongly influence the LO mode frequency while
the TO mode frequency may already be converged.
Since $\epsilon$ occurs in the denominator, overestimation
of $\epsilon$ in an odd sampling leads to underestimation of the
LO mode frequency.

Since a calculation of the full dispersion relation with a very dense
Monkhorst-Pack grid (and with the 5d electrons in the valence) was not 
feasible, we used a nested (7,3) grid. This is a 7$\times$7$\times$7
MP grid where, in addition, the cubic volume element around the point 
L is sampled by
a simple 3$\times$3$\times$3 grid. The high k-point density around L 
corresponds thus to the density in a uniform 21$\times$21$\times$21 grid 
and the solid line
in Fig.~\ref{samplingfig} demonstrates that we obtain the LO frequency in
very good agreement with the 20$\times$20$\times$20 and 19$\times$19$\times$19 samplings.

Due to the pronounced dip, the calculation
of the dispersion relation from a Fourier-interpolated 
dynamical matrix is therefore not feasible for the LO mode around
$\Gamma$. This is the reason, why in our dispersion relations
(Figs.~\ref{pbsdisdos},\ref{pbsedisdos},\ref{pbtedisdos}), we used interpolation for most
of the Brillouin zone but added point-by-point calculations
for the LO mode close to $\Gamma$.



\begin{thebibliography}{99}
\bibitem{gapnote}
In Ref.~\onlinecite{hummer}, Table IV, a similar overestimation of the PbTe
gap is observed, using the FP-APW+LO approach with the PBE functional.
\end{thebibliography}
\end{document}